\documentclass[aps,pra,twocolumn,groupedaddress,showpacs]{revtex4}

\usepackage{graphicx}
\usepackage{braket}

\begin{document}


\title{Dynamics of Low-Density Ultracold Rydberg Gases}


\author{J. O. Day}
\author{E. Brekke}
\author{T. G. Walker}
\email[]{tgwalker@wisc.edu}
\affiliation{Department of Physics, University of Wisconsin-Madison, Madison, WI 53706}


\date{\today}

\begin{abstract}
Population dynamics in weakly-excited clouds of ultracold $^{87}$Rb Rydberg atoms were studied by means of trap loss, fluorescence detection, and state dependent stimulated emission.  Rydberg atoms were excited to various $nl$ Rydberg states via continuous two-photon excitation from a magneto-optical trap.  A stimulated emission probe laser was then used to bring the Rydberg atoms down to the 6P$_{3/2}$ state, allowing state-dependent detection of the Rydberg atoms. Measurements of trap loss and fluorescent emission reveal information about the evolution of the Rydberg populations.  In particular, population in the initial Rydberg state quickly transfers to other Rydberg states by a non-collisional mechanism, likely superradiant emission.  The trap-loss measurements are consistent with black-body ionization as the dominant loss mechanism.
\end{abstract}

\maketitle


In the past 10 years it has become possible to study the interesting properties of dense clouds of ultracold Rydberg atoms, usually produced by laser excitation from  magneto-optical traps (MOTs) \cite{Mourachko98,Anderson98}.  A variety of interesting collisional phenomena \cite{Afrousheh04,Afrousheh06,Vogt06,Bohlouli07,Farooqi03,Li05,Singer05b,Amthor07,Carroll06} have been observed, including surprisingly rapid spontaneous conversion of dense ultracold Rydberg clouds into plasmas \cite{Robinson00,Li05,Walz-Flannigan04}.  It is clear that resonant \cite{Anderson98,Afrousheh04,Afrousheh06,Vogt06,Bohlouli07} and near resonant energy transfer collisions \cite{Li05,Singer05b,Amthor07} play important roles in the population dynamics of clouds of ultracold Rydberg atoms.

In addition, intriguing proposals to use the strong, long-range interactions between ultracold Rydberg atoms to perform conditional quantum manipulations \cite{Jaksch00} are driving a number of groups to study coherent manipulations with small numbers of atoms and small samples.  In addition, the concept of Rydberg blockade \cite{Lukin01} brings up the possibility of using  spatially confined samples of  atoms \cite{Sebby05,Yavuz06} for mesoscopic quantum manipulations.  Using such collective Rydberg excitations might allow novel quantum applications such as fast quantum gates \cite{Lukin01,Brion07}, single atom and directed single photon sources \cite{Saffman02}, and fast quantum state detection \cite{Saffman04}.

The ultracold Rydberg cloud experiments, with many Rydberg atoms excited at once, and the quantum manipulation experiments, with  only one Rydberg atom excited at a time, operate in very different regimes.
Still, it is interesting to note that in both cases
their success is reliant  on the strong interatomic forces between Rydberg atoms.  For quantum manipulations, one wants the Rydberg-Rydberg interactions to be so strong that multi-atom excitations are suppressed \cite{Walker08}.  A number of experiments \cite{Tong04,Singer04,Liebisch05,Heidemann07} report suppression signatures  suggesting that this regime should be achievable.  Most recently, collisional dephasing of coherent Rydberg Rabi flopping was observed for only two atoms confined in a dipole trap \cite{Johnson08}, and dephasing of coherent Rydberg excitation in a BEC was also observed\cite{Heidemann08}.

The experiment reported in this paper studies a transition regime between the two types of experiments.  We weakly excite ultracold Rydberg atoms in an un-blockaded regime and study their evolution using three tools:  trap loss,  spontaneous emission, and stimulated emission.  Using these diagnostics, we find that under our weak excitation conditions the probability is very small that the Rydberg atoms experience an inelastic collision sufficiently strong to leave the trap.  They predominantly return to the atomic ground state by spontaneous emission.  Using the stimulated emission diagnostic, we also observe rapid population transfer out of the initial Rydberg state--the initial Rydberg state is depopulated on a time scale substantially shorter than expected for transfer from black-body radiation or single-atom spontaneous emission.  The transfer rate slowly decreases with increasing principal quantum number, strongly suggesting that the transfer process is not collisional in nature.  We build upon the recent observation of superradiance by Wang {\it et al.} \cite{Wang07} and argue that superradiance is the likely mechanism for this transfer process.  Finally,  we show that trap loss rates are consistent with expectations from black-body ionization.  We conclude with implications of these results for other experiments.

\section{Trap Loss Studies\label{sec:traploss}}


In this section, we  describe observations of loss of atoms from the MOT due to Rydberg excitation.  Other groups \cite{Dutta01,Singer04}  have reported observations of Rydberg trap loss under various  conditions, without attempting to explain the mechanisms for trap loss, especially under the low excitation conditions reported here.  We find that trap loss rates are much less than excitation rates, showing that, once excited, the Rydberg atoms primarily decay  back to the ground state by spontaneous emission processes.

\begin{figure}
\includegraphics[scale=0.5]{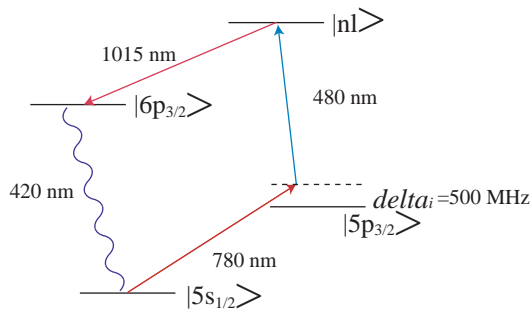}
\caption{Diagram of the Rydberg excitation scheme.}
\label{fig:elevel}
\end{figure}

We  excite magneto-optically trapped $^{87}$Rb $nl_j$ Rydberg atoms using continuous-wave 5S$_{1/2}\rightarrow$
5P$_{3/2}\rightarrow$nl$_j$   two-photon excitation. The 780 nm (20 mW) and 480 nm (10 mW) lasers used are tuned $\delta_i=500$
MHz above the intermediate 5P$_{3/2}$ F=3 state, thus avoiding direct excitation
of the 5P$_{3/2}$ state (Fig. \ref{fig:elevel}). Both excitation beams have spatial dimensions on the order of the $\sim$ 1 mm
size of the MOT. Typical two-photon excitation rates vary from 10 to 100/s depending on n level. These rates are deduced
from calculated atomic matrix elements, measured laser intensities, and the observed transition line-widths, and are averaged over the atomic and laser spatial distributions.  The effective
excitation rate is
${|\epsilon_2|^2/\Delta}$, where $\epsilon_2=\epsilon_r\epsilon_b/4\delta_i$ is the two-photon Rabi frequency,
$\epsilon_r$ and $\epsilon_b$ are  the single photon Rabi frequencies for the 780 nm and 480 nm lasers, and
$\Delta$ is the observed transition line-width which is typically 8 to 10 MHz for these scans as described below.  This
results in $\sim$10$^4$ Rydberg atoms at a relatively low density of 10$^7$/cm$^3$. This density is an order of magnitude or more smaller than the densities at which ultracold plasmas are formed \cite{Walz-Flannigan04,Li05}.  By changing the frequency of the 480
nm laser, we excite to  Rydberg states at 28D, 43D, 58D, and 30S; in most cases the D-state excitations are to the J=5/2
state. These particular states are chosen to provide a range of different spontaneous decay rates (which scale roughly as
$\sim n^{-3}$) and a wider range of Rydberg-Rydberg van der Waals  interactions $C_6 R^{-6}$, with $C_6(28,43,58)\approx (0.08,540,330)$ GHz $\mu$m$^6$\cite{Walker08}.

A very simple and effective way to observe excitation of Rydberg atoms in a MOT is via trap loss, with a resulting
decrease in the observed fluorescence from the MOT.  Processes such as black-body ionization,
photo-ionization by the applied lasers, and  inelastic collisions between Rydberg atoms and ground state atoms either ionize the atoms or give the
atoms sufficient ($>10$ K) kinetic energy to leave the trap even if they radiatively return to the ground state.  Radiative
processes such as photon emission or absorption,  elastic or inelastic collisions with cold electrons, and near-resonant  Rydberg-Rydberg collisions do not transfer
enough kinetic energy to the atoms to cause trap loss.  Trap loss is very sensitive; it is easy to observe a 0.1/s change
in the trap loss rate from the MOT.  With modest Rydberg excitation rates of 10/s, a 1\% inelastic  or
photo-ionization channel can easily be detected.

Since the excitation lasers are sufficiently off-resonance to cause negligible perturbations to the MOT trapping and
cooling processes, the loss due to production of Rydberg atoms affects the MOT only through a change in the ejection or loss rate from the trap.  If the MOT lasers produce a loading  rate $L$ and a loss rate $\Gamma_0$,  the steady-state number of trapped atoms is $N_{g0}=L/\Gamma_0$. Excitation of Rydberg atoms using lasers tuned to frequency $\nu$ adds a new loss rate $\Gamma_1$ which then changes the  number of atoms to $N_g(\nu)$.  The loss rate can then be found from
\begin{equation}
\Gamma_1(\nu)=\Gamma_0\left(\frac{N_{g0}}{N_g(\nu)}-1\right)
\end{equation}
This is operationally simpler than taking separate MOT loading transients at each frequency and determining the loss rate from the time constant.  We have checked that the two procedures give the same results.

\begin{figure}
\includegraphics[scale=0.6]{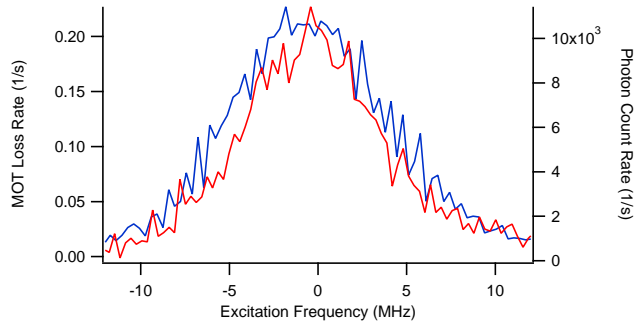}
\caption{Excitation beams are scanned over the Rydberg resonance at the 28D state. The peak excitation rate is 110/s. The figure shows the loss rate from the trap induced by the excitation beams as well as the photon count rate from the spontaneous emission cascade.}
\label{fig:2photon}
\end{figure}

Figure \ref{fig:2photon}
shows the results of sweeping
the  two-photon excitation frequency through a typical Rydberg resonance. Loss rates induced by the excitation lasers are around
0.2/s, a rate much smaller than the 110/s Rydberg excitation rate. We may infer from this that the probability of inelastic collisions occurring with sufficient energy transfer for the atoms to leave the trap is less than 1/500.  For the other states studied, we find 1/50 for 43D and 58D, and 1/1000 for the 30S state.

Noting that the trap loss probabilities are significantly higher for 43D and 58D as compared to 28D, a logical hypothesis  would be that inelastic collisions are responsible for the trap loss.  To this end, we compared the trap loss rates from 41D and 43D, and found little difference.  Since the 43D van der Waals interactions are anomalously large due to a near resonance in the 41P+45F and 43D+43D potentials \cite{Reinhard07,Bohlouli07,Walker08},  then very different results should be obtained when comparing 41D and 43D if collisions were responsible.  Thus we again find no evidence that resonant energy transfer collisions cause significant trap loss.

The following estimate supports these conclusions.  The inelastic collision rate should be a capture rate multiplied by a probability of energy transfer.  We estimate the capture rate as $\eta v\sigma $, where $\sigma=\pi R_0^2$, $C_6 R^{-6}=kT$, and $\eta v$ is the Rydberg atom flux.  We then deduce a capture rate of
\begin{equation}
\eta v\sigma =\eta v\pi \left(C_6/kT\right)^{1/3}\sim 200/s
\end{equation}
for an $n=43$ density of $\eta=10^{7}$/cm$^3$.  This rate, already small, is further reduced by the energy transfer probability.  For states like 43D, where the van der Waals interactions are repulsive at long range, there are no thermally accessible curve crossings and we expect the rate to be suppressed by at least a Boltzmann factor $e^{-\Delta E/kT}$, which even for 43D with small $\Delta E$ is a factor of $\sim 10^{-2}$.

The line-widths observed are significantly broader than the laser line-widths, and exceed what we expect from drifts of our reference cavity used to stabilize the 480 nm laser frequency.  This is consistent with observations by others \cite{Teo03,Singer04}.

Since the atoms are only rarely leaving the trap due to Rydberg-Rydberg collision or ionization, they must primarily return to the ground state by  emission of one or more photons.  In the next section we introduce a direct probe of the Rydberg state population that will give more information about the details of this process.

\section{Cascade Fluorescence}

 As a complementary diagnostic to trap loss observations, we use a photon counting module and a narrowband interference filter to detect the 420 nm decay photons from the 6P$_{3/2}$ state to the 5S$_{1/2}$ ground state. These photons are only observed coming from the MOT cloud under conditions of Rydberg excitation.  The fluorescent branching ratio to the 6P$_{3/2}$ from a high $n$S or $n$D Rydberg state is calculated to be about $b_r=0.15$, varying only slightly with principal quantum number.  (The predominant channel is emission to the 5P$_{3/2}$ state due to its much larger $\omega^3$ factor in the emission rate, which  easily compensates for a slightly smaller dipole matrix element. This channel is difficult to observe in the presence of strong MOT fluorescence at the same wavelength.) The probability of multiple photon cascade into the 6P$_{3/2}$ state is small because the long-wavelength photons required are disfavored. P and F Rydberg states predominantly cascade into 5S, 6S, and 4D levels, all of which lie energetically below the 6P state. Even for states slightly above the 6P, such as the 5D, the predominant decay channel is to states below the 6P. Thus the cascade fluorescence is likely a reliable probe of the S or D Rydberg state populations, and is relatively insensitive to P or F Rydberg states.  Figure \ref{fig:2photon}
 shows the observed cascade counts  as the excitation lasers are scanned across a Rydberg resonance.

The observed cascade signal has a background, mainly from dark counts, of around 250/s and reaches a peak signal of about 10000/s for the 28 D excitation. We can compare these observed rates with our expected rates from the calculated excitation rates and fluorescent branching ratios.
Accounting for  the finite collection solid angle $\Omega=3\times 10^{-3}$, $\eta=3.4$ \% detection efficiency of the photomultiplier tube, and a calculated 6P$_{3/2}$--5S$_{1/2}$ emission branching ratio $b_6=0.31$ we expect to observe a cascade count rate
\begin{equation}
c_6=R_2 N_g b_r b_6 \eta\Omega\frac{A_r}{A_r+A_{BB}}=18,000/{\rm s}
\end{equation}
at the peak of the 28D excitation resonance.  The  ratio of spontaneous to total decay rates accounts for the effects of black-body radiation causing radiative transitions to nearby Rydberg states.  These states are assumed, by the argument above, not to  result in detected cascade photons.
For the data seen in Fig. \ref{fig:2photon} we observe  10000/s, only 55\% of the expected count rate.  This suggests that roughly 1/2 of the Rydberg atoms are being transferred out of the excitation state by some other process.  The ratio of expected to detected cascade counts for the other excitation states (30S, 43D, and 58D) are (0.5, 0.6, 0.6). Again, by the arguments from the previous section, the explanation for this cannot be inelastic Rydberg-Rydberg collisions, which would either produce an extremely large trap loss rate in contradiction to observations, or would necessarily vary greatly with principal quantum number.

\section{ Stimulated Emission Probe}\label{sec:stateprobe}


For dipole blockade applications, one is particularly interested in the evolution of the blockaded Rydberg state.  Furthermore, in the particular applications of dipole blockade to single atom and single photon sources\cite{Saffman02}, stimulated emission is used to couple the blockaded Rydberg level to an intermediate atomic level.  Thus it is natural for us to pursue the development of a stimulated emission probe of Rydberg dynamics. The intensity dependence of the signals produced by the stimulated emission probe allows further information to be obtained about the population dynamics. Such a probe is non-destructive, and has  inherently high spectral resolution.  While being less general and less sensitive than field ionization, it does have the potential to be applied to a number of different states, subject to dipole selection rules.

As shown in Fig.~\ref{fig:elevel}, we apply a tunable diode laser in the range of
1013 to 1027 nm to perform stimulated emission probing of the Rydberg states produced by the two-photon excitation.  This ``state probe"  de-excites atoms from the Rydberg states to the 6P$_{3/2}$ state, which subsequently decays via a 420 nm photon to the 5S ground state. The 6P$_{3/2}$ state is either detected in the same manner as the cascade light mentioned above, or by a reduction in  trap loss (explained below). As with the excitation lasers, the spatial size of the state-probe laser is comparable to the MOT cloud size.  We deliver nearly 100 mW of light to the atoms, corresponding to stimulated emission rates from 5$\times$10$^5$ to 5$\times$10$^6$/s depending upon the Rydberg level involved. The natural decay rate of the 6P$_{3/2}$ level of 9.2$\times 10^6$/s is fast enough that population cannot accumulate there.  The state-probe laser is stabilized to a 300 MHz optical spectrum analyzer that is itself locked to a 780 nm Rb saturated absorption resonance.  It is tuned by changing the frequency of a double-passed acousto-optic modulator.

\begin{figure}
\includegraphics[scale=0.6]{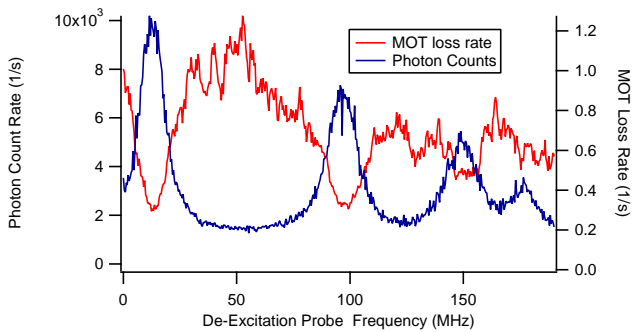}
\caption{Stimulated emission probe scan across the 6P$_{3/2}$ manifold. The cascade count background is around 1000 counts.}
\label{fig:scan}
\end{figure}

When the stimulated emission probe is on resonance, atoms are returned to the ground state more quickly than they would otherwise spontaneously radiate from
the long-lived Rydberg levels. This reduces the loss rate from the trap as the Rydberg atoms do not stay excited long enough for loss mechanisms such as black-body ionization and inelastic collisions to remove many of them from the trap. Keeping the excitation beams on resonance with the two-photon excitation, we can scan the frequency of the stimulated emission probe to observe the 6P$_{3/2}$ hyperfine manifold. Such a scan is shown in Fig.~\ref{fig:scan}.
 When tuned on resonance with the 6P$_{3/2}$ F=3 hyperfine state, the MOT loss rate is  reduced from 0.8/s to 0.2/s, with the amount of reduction depending on probe intensity.

\begin{figure}[t]
\includegraphics[scale=0.6]{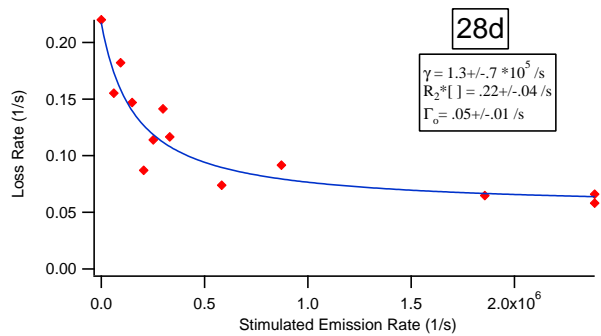}
\caption{Loss rate dependence on stimulated emission probe intensity for the 28D state, showing short residence times for the Rydberg state produced by two-photon excitation.}
\label{fig:loss}
\end{figure}

The state-probe laser produces a reduction in the trap loss rates.  An example of this for the 28D$_{5/2}$ Rydberg state is shown in Fig. \ref{fig:loss}.
 There the trap loss is shown as a function of the stimulated emission rate from the probe beam. The figure clearly demonstrates that by using a sufficiently high state-probe intensity the atoms can be returned to the ground state before ionization or inelastic collisions can occur,  thus reducing the loss rate of the MOT. The surprising feature of the data is that much higher intensities are needed than would be expected if the dominant population transfer from the Rydberg level were spontaneous decay or black-body transfer (rates of 4$\times$10$^4$/s and 2$\times$10$^4$/s, respectively)\cite{Gallagher}. The residence time of the Rydberg atoms is deduced to be nearly 10 $\mu$s.

\begin{figure}[h]
\includegraphics[scale=0.6]{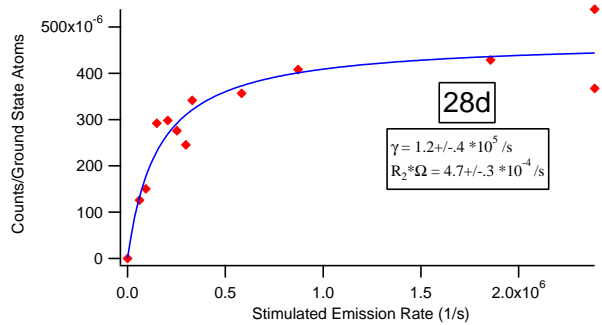}
\caption{Dependence of 6P$_{3/2}$ decay counts on stimulated emission probe intensity for the 28D state.}
\label{fig:counts}
\end{figure}

Alternatively, we can measure the state probe laser effects  by detecting the number of 6P$_{3/2}$ decay photons from stimulated emission from the Rydberg state. The count rate data, scaled by the number of ground state MOT atoms, is shown in Fig. \ref{fig:counts} for the 28D state. The cascade and background signal have been subtracted for this data, thus there are no counts when the stimulated emission rate is zero. As with the repletion data of Fig.~\ref{fig:loss}, the counts saturate at a stimulated emission rate of around $1\times$10$^5$/s, which implies that the other rates out of the excitation state must be on this order.

\begin{figure}
\includegraphics[scale=0.6]{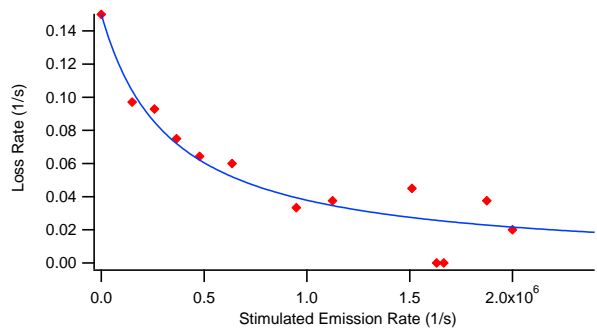}
\caption{Loss rate dependence on stimulated emission probe intensity for the 28D state, with the MOT magnetic field  switched off during excitation. }
\label{fig:pulsedB}
\end{figure}

From the trap loss or state probe data, the residence time of the Rydberg state varies from  about 10 $\mu$s at $n=28$ to about 50 $\mu$s at $n=58$.  Thus the population is being transferred out of the initial Rydberg state  faster than can be accounted for by spontaneous decay and black-body transfer (decay times of 25 $\mu$s and 50 $\mu$s at $n=28$, respectively). The transfer times increase with $n$, just the opposite dependence as would be expected for energy transfer collisions between Rydberg states.  These would be expected to decrease with $n$ due to rapid increase in van der Waals interaction strengths. This counter trend thus implies that Rydberg-Rydberg collisions are not the process responsible for the fast transfer out of the initial Rydberg state.

Interestingly, the 30S state has the highest transfer rate to other Rydberg states. This again implies that atom-atom interactions do not dominate the process, as the strengths of van der Waals  interactions tend to be larger for D states than for S states.

We have checked that the transfer rates increase with increasing excitation rate.  This confirms that some transfer process besides black-body radiation is occurring.

As can be seen from Fig. \ref{fig:loss},
there is a non-zero loss rate at high probe intensities where
 the stimulated emission laser should  fully deplete the original excitation state.  There are several possible explanations for this.  If there were a loss mechanism that was being enhanced by the state-probe laser, this would produce such a behavior.  However, processes such as photoionization or light-induced 6P-5S collisions can be estimated to be far too weak to account for this effect.  Another possibility has to do with Zeeman precession in the nd states causing population to accumulate in inaccessible magnetic sublevels. The magnetic field gradient used to confine the MOT atoms is large enough that precession between magnetic sublevels of the Rydberg atoms at the edges of the MOT  occurs at a rate of several MHz, which is on the order of the stimulated emission rate. The result of this precession is that a fraction of the atoms - up to 1/3 - move to a state that is inaccessible to the state-probe laser because of the dipole selection rules determined by the polarization of the state-probe laser. For linearly polarized state-probe light, $m=\pm5/2$ Zeeman levels cannot be excited to the 6P$_{3/2}$ state.  Thus population that accumulates in these levels cannot be de-excited by the state-probe laser.

This effect was verified by repeating the experiment with the MOT magnetic field being switched off for 10 ms intervals and only switching the Rydberg excitation lasers during the times the field was off.  This data, shown in Fig.~\ref{fig:pulsedB} for $n=28$ shows that Rydberg populations  at high state probe intensities are markedly reduced as compared to when the magnetic field is on.  The reduced signal-to-noise for this experiment made it possible to do this only for $n=28$.

\section{Model of Rydberg Population Dynamics}\label{sec:model}

In the previous sections, we have described the basic processes that are evidently at work under the conditions of our experiment, and their experimental signatures.  To further analyze the results, we present here a simplified model of the Rydberg dynamics and use it to extract the values of a few simple parameters from the data. Confirming the interpretation in the previous section,  we find that some process that does not cause trap loss nevertheless transfers population out of the excitation Rydberg state on a time scale short compared to spontaneous or black-body lifetimes.  The lengthening of this time scale with principal quantum number leads us to believe that it is not inelastic collisions between Rydberg atoms.

\begin{figure}
\includegraphics[scale=0.5]{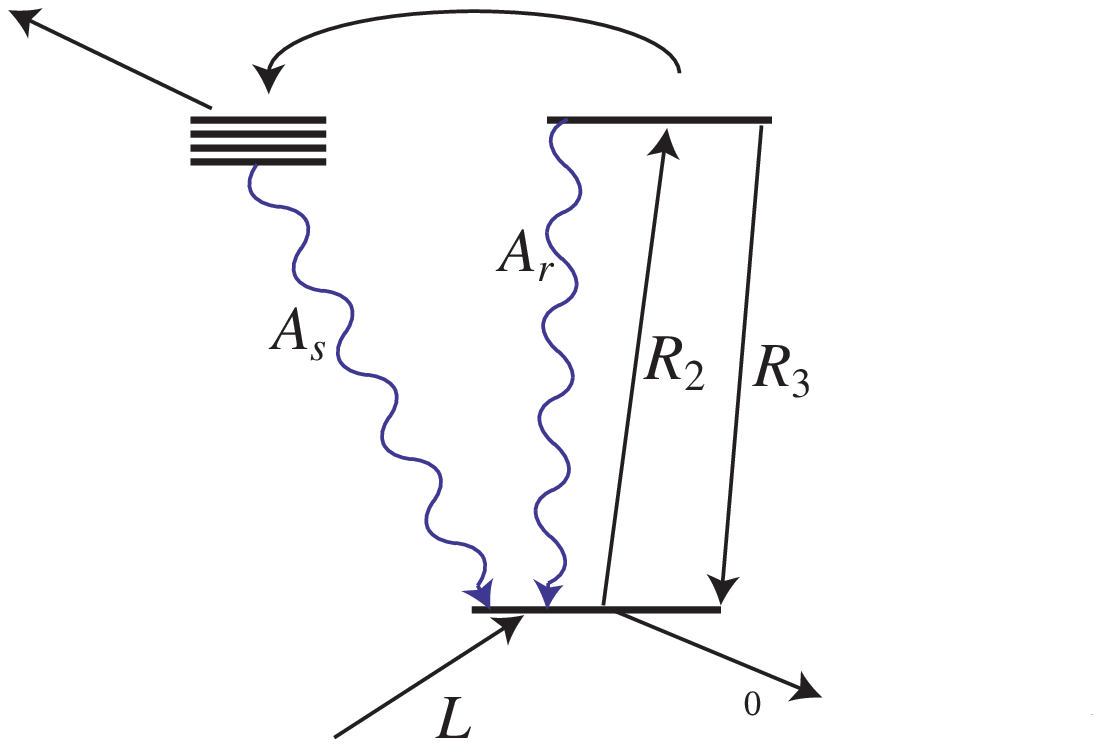}
\caption{Diagram of the simplified model of Rydberg population dynamics.}
\label{fig:model}
\end{figure}

The processes included in the model are illustrated in Fig.~\ref{fig:model}.  We  describe the system with a three state model: A ground state $\ket{g}$ with $N_g$ atoms, the excitation Rydberg state $\ket{r}$ with $N_r$ atoms, and an additional effective Rydberg state $\ket{s}$ that accounts for other states that are populated from state $\ket{r}$.

The dynamics of the excitation state $\ket{r}$ depend on laser excitation and de-excitation, spontaneous and black-body radiation, and transfer to the other Rydberg states $\ket{s}$. Population enters $\ket{r}$ by excitation from the ground state at a rate $R_2N_g$ where $R_2$ is calculated as described in Section~\ref{sec:traploss}.  Spontaneous decay to low-lying levels occurs at a rate $A_rN_r$.  Black-body radiation and other potential processes that transfer atoms to other Rydberg states occur at a rate $\gamma N_r$. There is also the possibility of trap loss (through ionization, for example) at a rate $\Gamma_r$ directly from state $\ket r$. Finally, de-excitation from the state-probe laser occurs at a rate $R_3N_r$.  Thus
\begin{equation}
\frac{dN_r}{dt}=R_2N_g-A_rN_r-R_3N_r-\gamma N_r -\Gamma_r N_r  \label{eqn:r} \\
\end{equation}
is the rate equation for the excitation state population.

The other Rydberg states are produced by collisional or radiative transfer from state $\ket{r}$ at the rate $\gamma N_r$ and have an effective radiative lifetime $A_s$.  We also assume that these states can cause trap loss at a rate $\Gamma_s$ due to black-body ionization and other collisional processes.  Thus they obey
\begin{equation}
 \frac{dN_s}{dt}=\gamma N_r-\Gamma_sN_s-A_sN_s \label{eqn:s}
\end{equation}
We are assuming that transfer from $\ket s$ back to $\ket r$ is unlikely.

In addition to the radiative de-excitation and excitation processes with the Rydberg levels, the ground state population $N_g$ is affected by MOT loading ($L$) and loss ($\Gamma_0$) processes that we assume are not materially changed when the Rydberg excitation lasers are on.  The resulting rate equation for the ground state population is
\begin{equation}
 \frac{dN_g}{dt}=L-\Gamma_0N_g-R_2N_g+(A_r+R_3)N_r+A_sN_s   \label{eqn:g}
\end{equation}
Plugging in the steady-state solutions to Eq.~\ref{eqn:r} and \ref{eqn:s} gives
\begin{equation}
 \frac{dN_g}{dt}=L-\Gamma_0N_g-\Gamma N_g   \label{eqn:g2}
\end{equation}
The loss-rate from the trap is increased by an amount
\begin{eqnarray}
\Gamma\approx\frac{R_2\gamma}{A_r+R_3+\gamma}\left[\frac{\Gamma_s}{A_s}+{\Gamma_r\over \gamma}\right] \label{eqn:loss}
\end{eqnarray}
This is a product of two factors.  The first essentially measures the excitation rate of Rydberg atoms, modified by the de-excitation due to the state-probe laser.  The second factor is the probability that the excited Rydberg atoms experience trap loss as opposed to radiatively decaying  back to the ground state.

The model similarly predicts the count rate of 420 nm photons produced by the state-probe laser:
 \begin{eqnarray}
\frac{I_3}{N_g}=\frac{R_3R_2\Omega\eta b_rb_6}{A_r+R_3+\gamma}
\end{eqnarray}
Thus  the state-probe-induced 420 nm count rate  can also be used to determine  the transfer rate $\gamma$, with the data  and fit for 28D shown in Fig. \ref{fig:counts}.

The results of fitting our experimental data to this model are summarized in Table~\ref{table:summary}.  We list there the deduced values of the primary unknowns $\gamma$ and $\Gamma_s$, as well as the assumed values for the input parameters $A_r$ and $A_s$.  We note that both trap loss and 420 nm count rates can be used to extract $\gamma$, an important internal consistency check.  The primary results are these:
\begin{itemize}
\item The atoms  transfer out of the excitation Rydberg state at a rate $\gamma$ that is substantially faster than spontaneous decay or black-body transfer rates.
\item The mechanism for population transfer, to an excellent approximation, does not cause trap loss.
\item The  population transfer rate decreases slowly with increasing principle quantum number, as opposed to the expected rapid increase if near-resonant energy transfer collisions were the relevant mechanism.
\item The probability of trap loss is very small; most Ryd\-berg excitations result in radiative repopulation of the ground state without trap loss.
\end{itemize}
These conclusions from fitting the experimental data to the model are consistent with the simplified analyses presented in Sections~\ref{sec:traploss}--\ref{sec:stateprobe}.  In the next two sections we will discuss their implications.

 \begin{table}[ht]
 \caption{A summary of  transfer rate data using both count rate
 and loss rate methods. As a basis for comparison, the black-body
 transfer rate A$_{BB}$, spontaneous emission rate A$_r$, the estimated mean spontaneous emission rate $A_s$ from other Rydberg states, and the inferred trap loss rate  $\Gamma_s$ are also included.
 All rates are in units of s$^{-1}$.}
    \centering
   \begin{tabular}{c c c c c c c}
   \hline\hline
   State & $\gamma$(counts) & $\gamma$(loss) & $A_{BB}$ & $A_r$ &
 $A_s$&$
   \Gamma_s$\\
   \hline
   28D& 1.2$\times$10$^5$& 1.3$\times$10$^5$& 2.6$\times$10$^4$& $4.1
   \times 10^4$& $3.1\times 10^4$& 265\\
   43D& 7.4$\times$10$^4$& 7.2$\times$10$^4$& 1.1$\times$10$^4$& 1.1$
   \times$10$^4$& $2.0\times 10^4$&602\\
   58D& 2.6$\times$10$^4$& 2.0$\times$10$^4$& 6.1$\times$10$^3$& 4.8$
   \times$10$^3$& $7.4\times 10^3$&433\\
   30S& 3.9$\times$10$^5$& 5.0$\times$10$^5$& 2.3$\times$10$^4$& 4.4$
   \times$10$^4$& $3.3\times10^4$&83\\
   \hline
   \end{tabular}
   \label{table:summary}
   \end{table}

\section{Superradiant Population Transfer}

We have argued above that the transfer mechanism responsible for population transfer out of the excitation state in a few microseconds cannot be due to near-resonant energy transfer collisions between excited Rydberg atoms.
L-changing collisions between Rydberg atoms and free electrons are another possible mechanism. Such collisions are unlikely to be energetic enough for the recoil to eject atoms from the trap and so would agree with much of the observed behavior. However, at our low excitation rates it is unlikely that electrons are present in sufficient quantities to cause the observed  state transfer.   With black-body ionization rates of 400/s (see Section~\ref{sec:bbi}) acting on a population of 10$^4$ Rydberg atoms, free electrons are produced at a rate of 4$\times$10$^6$/s. Free electrons are typically fast-moving and dissipate at a rate of $\sim$10$^4$/s \cite{Killian99}, giving an average free electron population of $\sim$400, which would not have a large impact on the 10$^4$ atoms in the excitation state. We estimate that there is insufficient ionization to cause an electron trap \cite{Killian99}. In addition, the rate of l-changing collisions should increase with principle quantum number, in contrast with our observations.

Having argued against collisional phenomena being responsible for Rydberg energy transfer, we need a radiative mechanism to explain our results.
Recently, Wang {\it et al.} \cite{Wang07} observed superradiance in the measured lifetimes of Rydberg states. That superradiance could play an important role in Rydberg population dynamics can be understood by the following arguments.  For principal quantum numbers $>20$, the size of the MOT is less than the wavelength for radiative emission from the excitation state to nearby dipole-allowed states.  If there are $N$ atoms initially in a particular Rydberg state, the collective dipole moment is enhanced by a factor of $N$ (assuming the atom cloud is much smaller than the wavelength of the emitted light).  The emission rate is enhanced by a factor of $N^2$, or by a factor of $N$ on a per-atom basis. Since $N$ is on the order of 10$^4$ for our experiment, the superradiant decay rate will be on the order of
\begin{equation}{2\over 3}{\omega^3 N d^2\over \hbar c^3}\sim {4N\over 3n^5}\alpha^3{Ry\over \hbar}\label{superest}
\end{equation}
which is $3\times 10^5$/s for 10$^4$ $n=50$ Rydberg atoms.  The low emission frequency is compensated for by the large number of cooperatively radiating atoms.  Since the spontaneous decay rate is proportional to $n^{-3}$,  the relative importance of superradiance and spontaneous decay goes only as $n^{-2}$, a factor of only 4 for this experiment.

Wang {\it et al.} \cite{Wang07} developed a sophisticated theory of superradiance with application to cold Rydberg gases.  Here we present a simplified model of superradiance to use as a interpretive guide.  We base this model on Dicke's original work \cite{Dicke54} as elucidated by Gross and Haroche \cite{Gross82} and Rehler and Eberly \cite{Rehler71}.

In considering the superradiant decay of an initial state $\ket{e}$ to a lower energy state $\ket{l}$, the Dicke approach introduces an effective collective spin state of the 2N-level system as $\ket{JM}$, with $N_e=J+M$ atoms in state $\ket{e}$ and $N_l=J-M$ atoms in state $\ket{l}$.  The radiation rate is found to be $\Gamma_{el}(J(J+1)-M(M-1))=\Gamma_{el}N_e(N_l+1)$.    To extend the two-level case to our multi-level case, we assume that we can model superradiance with a set of rate equations
\begin{equation}
{dN_e\over dt}=-\sum_{l<e}\Gamma_{el}N_e(N_l+1)+\sum_{l'>e}\Gamma_{l'e}N_{l'}(N_e+1)
\label{superrate}\end{equation}  This model reproduces the key features of superradiance:
a large initial inversion radiates at $\Gamma_{el}$ (per atom) at first, then as the inversion is reduced the rate accelerates to a maximum of $N\Gamma'_{el}/4$, occurring over a time $(\ln N)/(\Gamma_{el}N)\ll 1/\Gamma_{el}$ \cite{Rehler71}.  As an additional check on the model, we have simulated the experiment of Ref.~\cite{Gounand79} (which was performed at high temperatures and much smaller $n$) and our model reproduces the dominant features of the data shown there.

The emission rates $\Gamma_{el}$ are the rates for spontaneous emission multiplied by a cooperativity parameter $C_{el}$:
\begin{equation}
\Gamma_{el}=C_{el}{2e^2\omega_{el}^3\over  m c^3}{(2J_l+1)\over (2J_e+1)}f_{el}
\end{equation}
where the $f_{el}$ are the calculated absorption oscillator strengths.
For a system of atoms whose spatial extent is on the order of the wavelength of the transition, superradiance will occur at a reduced rate. The wavelength of the transition to the nearest lower lying Rydberg state varies from 0.17 cm for 30S-29P to 2.8 cm for the 58D-59P transition. For low $n$, this is quite close to the 1 mm spatial extent of the MOT, and lower levels will have even shorter transition wavelengths.
Following Ref.~\cite{Rehler71},  the cooperativity parameter for a uniform density system of N atoms in a volume V radiating in direction $\hat{k}$ is (in the $N\gg1$ limit)
\begin{eqnarray}
C_{el}&=&\frac{1}{V^2}\int d^3x\int d^3x^{'}\,e^{i(\vec{k}-\vec{k}_1)\cdot(\vec{x}-\vec{x}^{'} )}\\
&=&\frac{9(\sin (k_{el}R)-k_{el}R \cos (k_{el}R))^2}{(k_{el}R)^6},\end{eqnarray}
for a spherical uniform density cloud, where $kR$ is the product of the wavenumber and the radius of the atomic sample.  The cooperativity parameter is 1 for  $R\ll \lambda$ and decreases to $0$ for $R\gg\lambda$.  In practice, the cooperativity parameter becomes small around $n=20$ for a 1 mm MOT.

To account for superradiance, we replace the level $\ket s$ of the model of Section~\ref{sec:model} with a set of levels near in energy to $\ket r$ that are coupled to each other and to $\ket r$ by  black-body radiation and by superradiance from Eq.~\ref{superrate}.  We find steady-state solutions of the resulting non-linear equations and from them deduce the effective Rydberg-Rydberg transfer rate from Eq.~\ref{superrate} with $e=r$.

Table II compares the observed Rydberg transfer rates and those predicted by our simulation.  These rates do not drop off as quickly as would be expected from Equation~\ref{superest}.  This is a result of the spatial factor approaching unity for the higher n-levels, balancing out the decrease in natural emission rate.  Additionally as a result of this increasing spatial factor, our simulation indicates that the atoms are transferred predominantly to f-states for the 43d and 58d levels.  This transfer to f-states could explain the higher total loss rate from the trap for these levels.  We emphasize that there are no adjustable parameters in our simplified superradiance model, and yet it naturally predicts the order of magnitude of the Rydberg-Rydberg transfer rates.  A more sophisticated model would be expected to explain the variation seen.

\begin{table}[ht]
\caption{A comparison of Rydberg-Rydberg transfer rates deduced by comparing measurements deduced from the model of Section~\protect\ref{sec:model} with those predicted by the superradiance model.  All units are s$^{-1}$.}
\centering
 \begin{tabular}{c c c}
 \hline\hline
 State & $\gamma$(calculated) & $\gamma$(expt) \\[0.5ex]
 \hline
 28D & 1.7$\times$10$^5$ & 1.3$\times$10$^5$ \\
 43D & 2.4$\times$10$^5$ & 7.4$\times$10$^4$ \\
 58D & 1.2$\times$10$^5$ & 2.0$\times$10$^4$ \\
 30S & 2.2$\times$10$^5$ & 5.0$\times$10$^5$ \\
 \hline
 \end{tabular}
 \label{table:comp}
 \end{table}

\section{Black-Body Ionization\label{sec:bbi}}

So far we have not discussed the actual mechanism for trap loss.  The deduced trap loss rates are quite modest, typically 400/s on a per Rydberg atom basis.  This is close to what would be expected from black-body ionization.  Black-body ionization rates were recently calculated in Ref.~\cite{Beterov07} for the various Rydberg levels.  Assuming Rydberg $n$-level distributions from our superradiance model, we find expected photoionization rates from black-body radiation to be as shown in Table~\ref{table:bbi}.
For the D states, there is reasonable agreement between our deduced experimental rates and the predicted values,   suggesting that black-body ionization probably  composes a large portion of the total loss from the trap.
For the s-state, the predicted trap loss is greater than observed by about a factor of 3, for which we have no explanation.

 \begin{table}[ht]
 \caption{Comparison of inferred Rydberg trap loss rates with black-body ionization
 rates.  $\Gamma_{BBI}$ is the rate of
 black-body ionization for each of these states calculated from Ref.~\protect\cite{Beterov07}. All rates are in units of s$^{-1}$.}
   \centering
   \begin{tabular}{c c c}
   \hline\hline
   State& $\Gamma_{s\,calc}$&$\Gamma_{BBI}$ \\ [0.5ex]
   \hline
   28D  & 212 & 322 \\
   43D  & 470 & 720 \\
   58D  & 329 & 457 \\
   30S  & 77 & 265 \\
   \hline
   \end{tabular}
   \label{table:bbi}
   \end{table}

\section{ Discussion}

In this paper, we have presented results on the dynamics of low-density ultracold Rydberg clouds excited using two-photon absorption from a MOT.  By looking at trap loss, radiative cascade, and state-selective stimulated emission we find that the probability of the Rydberg atoms undergoing collisional loss before decaying back to the ground state is small.  Additionally, the Rydberg state produced by the two-photon excitation is depleted by some mechanism on a time scale significantly shorter than can be explained by black-body transfer or single-atom radiative decay.  This time scale increases with increasing principal quantum number, in contradiction to expectations if
 inelastic Rydberg-Rydberg collisions were responsible for the rapid state transfer.   On the other hand, estimates of collective superradiant light emission and a simplified model thereof suggest that this mechanism can explain the rapid population transfer observed in this experiment.  The overall trap loss rates are consistent with expectations from black-body ionization.

Before the experiment of Wang {\it et al.}\cite{Wang07} the effects of superradiance were not appreciated for the population dynamics of high density ultracold Rydberg atoms.  Superradiance can happen on very rapid time scales, especially under strong Rydberg excitation as achieved in a number of previous experiments.  In fact, superradiance can easily occur on sub-microsecond time scales.  A key consequence of superradiance is that it populates Rydberg states lying energetically below the state being excited by the laser.  Indeed, in the experiments of Ref.~\cite{Walz-Flannigan04} and Ref.~\cite{Li04}, population clearly moves to  lower lying Rydberg levels on a fast time scale, consistent with the hypothesis of superradiant transfer.

An additional consequence of fast superradiant population transfer is that it provides a mechanism for rapid population of states of neighboring orbital angular momentum $l$ from the excitation state.  Pairs of atoms with $\delta l=\pm 1$ interact at long range via the $R^{-3}$ resonant dipole-dipole interaction, not the usual $R^{-6}$ van der Waals interaction. The much stronger collision interactions between these atoms may explain the very rapid time scales for plasma formation in a number of experiments where resonant dipole-dipole interactions were not purposely produced using external fields.

\begin{acknowledgments}
J. Strabley contributed to the early stages of this work.  We appreciate helpful discussions with M. Saffman and D. Yavuz.  This work was supported by the National Science Foundation and NASA.
\end{acknowledgments}

\bibliography{lasercooling}
\end{document}